\documentclass[runningheads,review,anonymous]{llncs}
\usepackage[T1]{fontenc}
\usepackage{graphicx}
\usepackage{booktabs}
\usepackage{algorithm} 
\usepackage{algpseudocode}
\usepackage{amsmath}

\algnewcommand{\algorithmicand}{\textbf{ and }}
\algnewcommand{\algorithmicor}{\textbf{ or }}
\algnewcommand{\OR}{\algorithmicor}
\algnewcommand{\AND}{\algorithmicand}
\usepackage{amssymb}
\usepackage{listings, xcolor}
\definecolor{javapurple}{rgb}{0.5,0,0.35} 
\definecolor{linenumbergray}{rgb}{0.5,0.5,0.5}
\lstdefinestyle{Java-github}{
        basicstyle=\ttfamily\tiny,
        language=Java,
        commentstyle=\color{linenumbergray},
        stringstyle=\color{javapurple},
        keywordstyle=\color{red},
        morekeywords={@Test},
        morecomment=[s][\color{linenumbergray}]{/**}{*/},
        numbers=left,
        numberstyle=\tiny\color{linenumbergray},
        numbersep=2.5pt,
        xleftmargin=1em,
        moredelim=**[is][\color{javapurple}]{@h@}{@h@},
        morecomment=[f][{\btHL[fill=gitdel]}]-,
        morecomment=[f][{\btHL[fill=gitadd]}]+,
        breaklines = true,
}
\lstdefinestyle{prompt}{
     basicstyle=\ttfamily\scriptsize,
     language=Html,
     commentstyle=\color{linenumbergray},
     stringstyle=\color{javapurple},
     keywordstyle=\color{red},
     morekeywords={@Test},
     morecomment=[s][\color{linenumbergray}]{/**}{*/},
     numbers=left,
     numberstyle=\tiny\color{linenumbergray},
     numbersep=2.5pt,
     xleftmargin=1em,
     moredelim=**[is][\color{javapurple}]{@h@}{@h@},
     morecomment=[f][{\btHL[fill=gitdel]}]-,
     morecomment=[f][{\btHL[fill=gitadd]}]+,
     breaklines = true, 
     escapeinside={(*@}{@*)}
}
\begin{document}
\title{APRIL: API Synthesis with Automatic Prompt Optimization and Reinforcement Learning}
%
%
\author{Hua Zhong\inst{1} \and
Shan Jiang\inst{1} \and
Sarfraz Khurshid\inst{1}}
\authorrunning{Anon. Author et al.}
\titlerunning{APRIL: API Synthesis with APO and Reinforcement Learning}
%
\institute{The University of Texas at Austin, Austin TX 78712, USA
\email{\{hzhong,shanjiang\}@utexas.edu, kuhrshid@ece.utexas.edu}}
%
\maketitle              

\newcommand{\NumTasks}{81}
\newcommand{\NumTasksCompletedOurApproach}{133}

\begin{abstract}

APIs are central to modern software development, yet composing new APIs from large libraries is difficult due to the exponential search space; traditional component-based synthesis relies on costly exploration and hand-crafted specifications. While large language models (LLMs) can generate implementations from natural language, hallucinations and limited access to up-to-date contextual information often yield incorrect code. In this paper, we present APRIL, an approach that combines LLM-based synthesis with Automatic Prompt Optimization (APO) and Reinforcement Learning from Verifiable Rewards (RLVR): APO iteratively refines prompts for a frozen model, while RLVR fine-tunes the policy toward functional correctness, producing an efficient synthesis pipeline. Evaluated on 81 real-world APIs from widely used scientific Python libraries and benchmarked against instruction-tuned but unfine-tuned LLMs guided by expert prompts, APRIL achieves substantial improvements. These results indicate that integrating APO and RLVR provides a robust, scalable path for component-based API synthesis in large libraries.

\end{abstract}
\section{Introduction}
\label{sec:intro}

The increasing complexity of modern software systems has made correct API usage and library integration a significant challenge for developers \cite{lamothe2021systematic}. Even experienced programmers often spend considerable time identifying relevant functions within a library and must frequently combine these with complex control structures to achieve desired functionalities \cite{jungloid,frangel}. This complexity complicates program synthesis, requiring both knowledge of library components and their effective integration into coherent code. Correct API synthesis need to capture intended behavior, which is often informal and difficult to express as formal specifications. While formal methods such as logical constraints \cite{10.1145/1993316.1993506,proveri} and executable specifications \cite{mimic,oguide} exist, synthesizing APIs from input-output examples has become a more user-friendly and accessible alternative \cite{frangel}. This approach allows developers—including non-programmers—to specify desired functionality using examples, bypassing the need for detailed technical specifications \cite{stringio}.



Prior work on program synthesis has largely relied on domain-specific knowledge—e.g., string manipulation \cite{icmlpbe,stringio} or data transformations \cite{syndstrans,syntable}—or on expert-crafted domain-specific languages (DSLs) \cite{testdriven}. Domain-agnostic techniques such as loop-free synthesis \cite{loopfree} and oracle-guided synthesis \cite{oguide} instead require formal specifications of components; specifying precise behavior for every library primitive is costly and often infeasible, particularly when intended behavior is informal or tacit. Although these methods solve various tasks, they typically target single basic blocks and scale poorly to multi-block programs with loops, recursion, or complex guards, due to combinatorial growth in method sequences. To mitigate this, SyPet \cite{sypet}, EdSynth \cite{Edsynth}, and FrAngel \cite{frangel} favor black-box execution over formal semantics, reducing dependence on DSLs and specifications. SyPet, however, struggles to synthesize rich control flow; FrAngel introduces angelic conditions with iterative refinement to construct more complex programs; and EdSynth leverages program sketches to similar effect. Live-execution approaches such as LooPy \cite{loopy} involve programmers as oracles, improving interactivity but reintroducing human dependence and limiting full automation in API synthesis.


In contrast, large language models have recently demonstrated remarkable proficiency in natural language understanding and code synthesis tasks\cite{nam2024using,jiang2024generating}. Trained on large datasets, including extensive code repositories and documentation, LLMs possess an implicit understanding of programming syntax and semantics. These models can align user inputs with their pre-existing knowledge and even address tasks that are uncommon in their training data \cite{schaeffer2024emergent}. Unlike traditional synthesis methods, which often rely on domain-specific knowledge\cite{sypet} or predefined formal specifications, LLMs excel at managing complex, open-ended programming tasks. They can synthesis APIs based on function signature and other part of the module, eliminating the need for explicit specifications or DSLs. Furthermore, LLMs handle intricate control structures such as loops and conditionals with ease, making them highly versatile and efficient. By directly interpreting developer intent through function signatures and I/O examples, LLMs reduce the burden on developers, enabling rapid and adaptive solutions for diverse functionalities.

However, industry experience indicates that large language models (LLMs) continue to exhibit limitations, such as hallucinations \cite{nikolov2025google,jiang2025cascade}. Furthermore, there is a lack of research on systematically evaluating the ability of LLMs to synthesize code for deep learning libraries. In this paper, we evaluate the effectiveness of Automatic Prompt Optimization (APO) \cite{pryzantetal2023automatic} and Reinforcement Learning from Verifiable Rewards (RLVR) \cite{lambert2025tulu} approaches for API synthesis in scientific Python libraries. APO refers to methods that autonomously refine prompts to improve task performance in large language models. APO techniques offer key advantages: they enhance model outputs without requiring access to model parameters, systematically explore the space of possible prompts, and produce improvements that remain interpretable to humans. In addition, RLVR optimizes the policy using rewards derived from objective, programmatically checkable signals rather than human preference models. APO refines prompts, instructions, and in-context examples to bias a frozen LLM toward producing correct implementation. RLVR prioritizes functional correctness and safe component composition by grounding feedback in unit-test outcomes and static-analysis diagnostics, thereby improving pass rates and reliability. This allows the model to internalize higher-level conventions and reliably apply them in synthesis tasks.

While both approaches have shown promise in various machine learning contexts, there has been no systematic comparison of APO and RLVR for the synthesis of new APIs in scientific Python libraries. Prior research has largely focused on API invocation accuracy or small-scale code completion tasks, rather than on the end-to-end generation of maintainable, idiomatic, and robust API implementations. The reliability stakes are higher here: a poorly synthesized implementation, once adopted, can ossify bad patterns across a codebase and become expensive to replace. In this paper, we present APRIL, the first controlled, reliability-centric evaluation of APO v.s. RLVR for synthesizing new API implementations in scientific Python libraries.

This paper makes the following contributions:
\vspace{-4pt}
\begin{itemize}
    \item {\bf Approach}. We present a novel approach to API synthesis using LLMs by integrating Automatic Prompt Optimization (APO) and Reinforcement Learning from Verifiable Rewards (RLVR). By employing APO to iteratively refine prompts and leveraging RLVR for efficient model fine-tuning, our method reduces reliance on exhaustive search and manual component specification.
    \item {\bf Evaluation}. We conduct an experimental evaluation of our approach on 81 programming tasks sourced from widely used scientific Python libraries, including NumPy, SciPy, and scikit-learn. We benchmark our method against state-of-the-art, un-finetuned large language models (LLMs) that rely on human-curated prompts. Experimental results demonstrate that our approach achieves a success rate exceeding 93.8\% in synthesizing correct, test-passing APIs, substantially outperforming baseline LLMs in terms of accuracy.
    \item {\bf Test case generation}. To obtain the verification oracles required for APO training and RLVR fine-tuning, we employ the Gemini CLI coding agent~\cite{geminicli} to automatically synthesize test suites for scientific Python APIs in our training set. We additionally evaluate Gemini CLI’s test generation capability, assessing the extent to which the produced suites are comprehensive and semantically meaningful for validating API behavior. This analysis underscores Gemini CLI’s promise as an enabling component for automated API synthesis pipelines and rigorous evaluation workflows.
\end{itemize}

\section{Problem Definition}
\label{sec:problem}
This paper proposes a novel methodology for component-based API synthesis, which leverages large language models (LLMs) to generate an API from a set of base components and input-output examples. Similar to traditional techniques, the synthesis problem in our proposed methodology formally requires the following inputs:
\begin{itemize}
    \item A test oracle that, given a candidate API, returns a Boolean verdict (true/false) indicating whether the candidate satisfies the intended task specification. Concretely, we instantiate the validation oracle using a comprehensive test suite generated via an AI-agent-based methodology. The procedure for constructing this test suite is detailed in Section~\ref{sec:sec31}.
    \item A method signature that precisely specifies the API’s parameter and return types, as well as its invocation modality (in Python, an instance method, class method, static method, or module-level method).
    \item A library of components from which the synthesizer assembles candidate implementations of the target API. Specifically, we provide the library name and the module path of the API’s intended location to the LLM as contextual metadata, enabling it to determine the components required for synthesis.
    \item A set of examples that, for any input within the API’s valid domain, returns the reference output of the target API. The example set may be a subset of the validation tests or constitute an independent collection of tests.
\end{itemize}
\section{API Synthesis Methodology}
\label{sec:approach}

\begin{figure*}[h]
    \centering
    \includegraphics[width=1.0\textwidth,trim={6cm 8cm 14cm 6cm},clip]{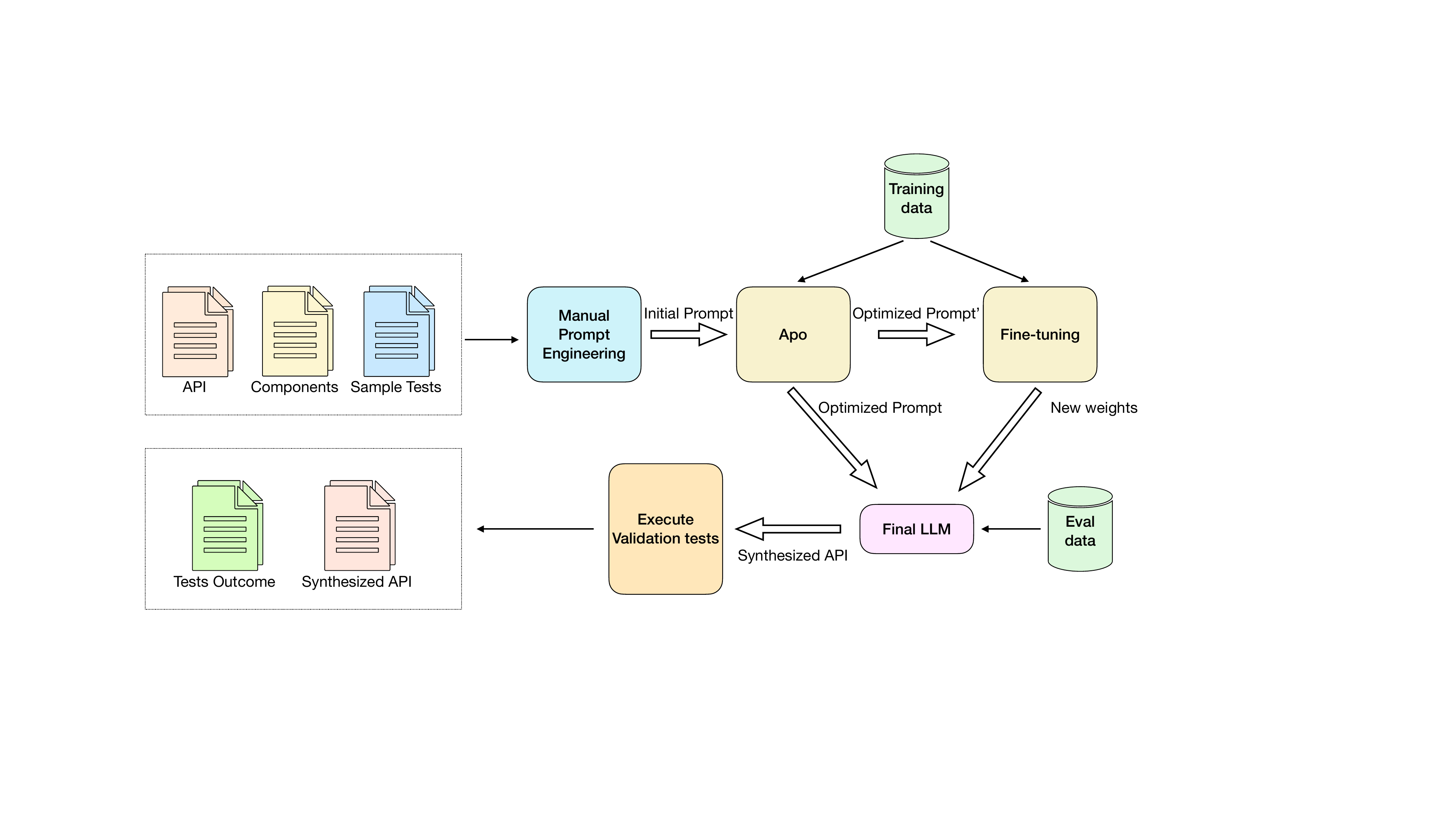}
    \caption{Workflow of APRIL}
    \label{fig:figure1}
\end{figure*}

In this section, we illustrate the architecture of APRIL, and break it down into different components. The overall workflow of APRIL is shown in Fig.~\ref{fig:figure1}. Our study focuses on synthesizing Scientific Python APIs. As mentioned in Section~\ref{sec:problem}, validating the correctness of the synthesized API requires a test-based validation oracle. We construct this oracle using a methodology powered by Gemini-cli, which automatically generates the requisite test suite. APRIL further requires a component library, the target method signature, and a set of tests. These inputs are used to construct the API-synthesis prompt, after which we employ LLMs to directly generate candidate method implementations. To improve synthesis effectiveness, we apply Automatic Prompt Optimization (APO) and RLVR (Reinforce Learning from Verifiable Rewards) fine-tuning, guiding the LLMs toward the desired outputs. The details of our prompt-engineering methodology are presented in the following sections.

\begin{algorithm} [t]
    \caption{Validation Tests Generation}
    \label{validationoraclegeneration}
    \begin{algorithmic}[1] 
        \Procedure{GenerateValidationTests}{$ds$, $ri$}
            \State $feedback_{t} \gets \varnothing$
            \State $feedback_{c} \gets \varnothing$
            \While{$true$}
                \State $T_c \gets genTests(ds, ri, feedback_{t}, feedback_{c})$
                \If{$testPass(T_c, ri) \AND isGoodQuality(ds, ri, T_c)$}
                    \State \textbf{return} $T_c$
                \Else
                    \State $feedback_{t} \gets testPass(T_c, ri)$
                    \State $feedback_{c} \gets isGoodQuality(ds, ri, T_c)$
                \EndIf
            \EndWhile
        \EndProcedure
    \end{algorithmic}
\end{algorithm}


\subsection{Validation Oracle Generation}
\label{sec:sec31}
For each target API, we construct a comprehensive test suite that serves as its validation oracle. Because this test suite is only used to validate the synthesized API and is not an input to the synthesizer, we assume access to the API’s docstrings $ds$ (natural-language documentation embedded in the code) and a correct reference implementation $ri$. These artifacts are supplied to an AI agent, which generates the corresponding tests. Concretely, given $ds$ and $ri$, we instruct the agent to produce a candidate test suite $T_c$ intended to validate the API. The agent then executes $T_c$ against $ri$ and reports the results; if any test fails, the failure information is fed back to the agent, which regenerates $T_c$. Coinciding with this, we submit $ds$, $ri$, and $T_c$ to a second LLM acting as an evaluator to assess the quality of $T_c$ with respect to comprehensiveness and test coverage (i.e., breadth of input scenarios and exercised behaviors). If the evaluator deems the quality insufficient, its feedback is returned to the AI agent, which revises $T_c$ accordingly. We output the revised $T_c$ as the final validation oracle once all tests pass and the quality criteria are satisfied. The algorithmic details of this test-generation procedure are presented in Algorithm \ref{validationoraclegeneration}.

\begin{figure}
    \centering
        \begin{lstlisting}[style = prompt]
### Implement a python method inside a python module. 

**Task:**

Imagine you are a Python developer specialized in machine learning and scientific libraries. You will be given a python method signature, the module it belongs to, the library it belongs to, and a set of test cases. The method can be a instance method of a class or a module method. Your task is to implement the method using the dependencies from the module and the library to pass the test cases.

**Output Format:**

Put your ouput inside the <output_api_implementations> xml element, and output only the implemented method.
<output_api_implementations> [your method implementation here] </output_api_implementations>

**Crucial Information:**

Carefully analyze the provided test cases. Each test case represents a specific input and the expected output of the method you are implementing. Your implementation *must* satisfy all given test cases. Consider these test cases as concrete examples of how the method should behave. Aim to re-use functions from the specified library wherever possible.

**Here are some examples:**
Method signature: ****
Module: ****
Library: ****
Test cases: ****
Output:
<output_api_implementations>
****
</output_api_implementations>

****

Method signature: ****
Module: ****
Library: ****
Test cases: ****
Output:
<output_api_implementations>
****
</output_api_implementations>

Method signature: {api_signature_rlvr_apo}
Module: {module_rlvr_apo}
Library: {library_rlvr_apo}
Test cases: {tests_rlvr_apo}
Output:
        \end{lstlisting}
    \caption{Manually engineered initial prompt}
    \label{fig:figure2}
\end{figure}

\subsection{Initial Prompt Construction}
\label{sec:sec32}
We define the initial prompt as the API-synthesis prompt used both as the input to the baseline model and as the seed prompt for Automatic Prompt Optimization (APO). We construct this prompt manually using established prompt-engineering techniques, including: (i) role conditioning (“assistant creation”), which assigns an explicit persona to the LLM to situate the task context and constrain response style; (ii) chain-of-thought prompting, which encourages decomposition of the synthesis task into smaller subproblems and the production of intermediate reasoning steps; and (iii) few-shot learning, which supplies a small set of input–output exemplars to prime the LLM’s behavior through in-context learning. The resulting initial prompt is shown in Fig.~\ref{fig:figure2}.

\subsection{Automatic Prompt Optimization}
\label{sec:sec33}
Automatic Prompt Optimization (APO) is a model-agnostic procedure that operates with minimal prerequisites: a small training dataset, an initial prompt, and API access to an LLM. The algorithm processes mini-batches of training examples to elicit “gradient”-like signals: textual critiques that diagnose limitations of the current prompt. It then uses these signals to propose and apply prompt edits. In practice, APO performs iterative prompt refinement analogous to gradient-based updates.

\begin{figure*}[h]
    \centering
    \includegraphics[width=0.86\textwidth,clip]{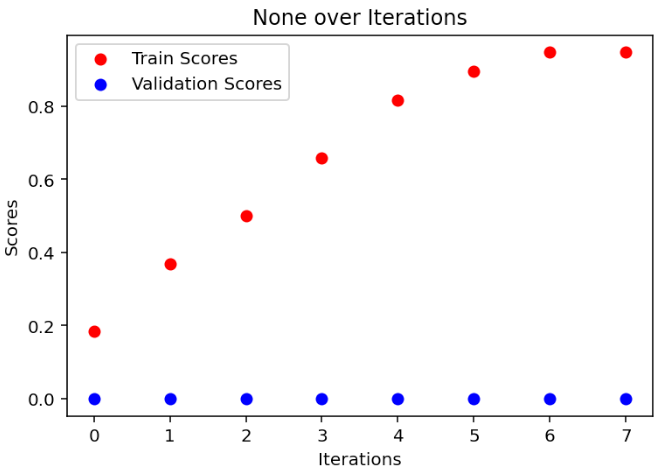}
    \caption{Example of iterative improvement in the discriminator score ($DS$) between APO search iterations}
    \label{fig:figure3}
\end{figure*}

In our approach, we employ the manually engineered prompt from Section~\ref{sec:sec33} as the initial prompt $P_0$ and use a subset of the benchmark dataset as the training data set $\mathcal{A}$ of $n$ training tasks. Within our APO framework, text-gradient generation serves as the primary mechanism.

To illustrate, consider a prompt candidate $P$ and its generated solution ${s_a}, {a\in\mathcal{A}}$ at the current iteration. We provide $P$, $s_a$, and $\mathcal{T}a \in \mathcal{T}$ (where $\mathcal{T}$ denotes the validation test suite) to a discriminator prompt that instructs a classification LLM $llm_c$ to execute the tests. We first request a binary verdict (fail/pass) from $llm_{c}$ and use it to compute a bad-generation penalty $p_a\in{0,1}$ from this result. If any test fails, we then request a textual critique $e_a$ from $llm_{c}$ describing the observed errors produced by $p_a$. After processing all $a \in A$ in the current iteration, we compute the discriminator score $DS = 1- \frac{1}{n}\sum_{i=1}^{n} p_i$ and form the iteration’s ``text gradient'' by concatenating the textual critiques ${e_i}$. Figure~\ref{fig:figure3} illustrates the iterative improvement of the discriminator score ($DS(P)$) over successive APO iterations. We then follow the standard APO procedure to propose edits to the current prompt and perform beam search over the resulting candidates, using $DS(P)$ to sort prompts at each level of the search tree. The candidate $P^\star$ with the highest $DS(P)$ in the final iteration is selected as the output of our APO algorithm.

\subsection{RLVR Fine-tuning}
\label{sec:sec34}
Loss function adopted by classic fine-tuning on text generation tasks is typically the Negative Log-Likelihood (NLL) loss. The NLL loss quantifies the negative log-probability that the model assigns to the ground-truth next token (i.e., the discrepancy between the predicted next-token distribution and the empirical distribution concentrated on the observed token). Minimizing NLL promotes higher probabilities for correct tokens and lower probabilities for incorrect tokens. In the context of program synthesis, however, NLL primarily captures surface-level similarity between the target and LLM-generated implementations and is insensitive to functional equivalence. This limitation is consequential, as distinct programs can be semantically interchangeable despite syntactic divergence.

Given our evaluation is driven by a verification test suite, we recast fine-tuning from next-token prediction to RLVR, where synthesized programs receive rewards directly based on test outcomes (functionally correct versus incorrect). 

To obtain training signals, we reuse the training dataset from Section~\ref{sec:sec33} and, for each API task x, condition the policy model $\pi_\theta$ on the task prompt to sample $K$ candidate implementations $S={s_1,\dots,s_K}$. Sampling uses stochastic decoding (temperature and top-$p$) with de-duplication of identical code strings.
Each $s_i$ is executed against the validation tests. We assign a binary reward $R(s_i)\in{0,1}$, with $R(s_i)=1$ iff all tests pass, and $0$ otherwise. Candidates that fail to build and execute receive $R(s_i)=0$. We optimize $\pi_\theta$ with Group Relative Policy Optimization (GRPO). For each group $S$, we compute a group-relative advantage by centering the reward of $s_i$ against the group mean, reducing variance and inducing competition. Updates use a PPO-style clipped surrogate with a KL penalty to a reference policy; $\pi_{\theta_{\text{old}}}$ is fixed during each update and periodically refreshed. Training runs over mini-batches for a few epochs with early stopping when group reward stabilizes.

\section{Evaluation}
\label{sec:evaluation}
In this section, we present a comprehensive experimental evaluation to assess the effectiveness of APRIL in scientific Python API synthesis tasks. The primary objective is to evaluate the practicality of the generated APIs and their capability to accurately implement the expected functionality. Specifically, we address the following research questions:

\begin{itemize}
    \item \textbf{RQ1: Effectiveness.} What proportion of APRIL generated APIs successfully build and pass the validation tests?
    \item \textbf{RQ2: Comparison.} What is the relative success rate of APRIL, augmented with APO and RLVR fine-tuning, compared with a baseline model employing a manually engineered prompt?
    \item \textbf{RQ3: Tests Generation.} How effective is the AI agent adopted in our study on tests generation? 
\end{itemize}

\subsection{Experimental Setup}
\label{sec:sec41}
To address the above research questions, we designed an experimental framework incorporating the following key components:

\subsubsection{Large Language Models}
We employ a customized instance of Gemini 2.0—an instruction-tuned large language model—as the synthesis model. The sampling temperature is set to 0.7 to balance diversity and determinism in generated code, a configuration empirically shown to be effective for code generation tasks \cite{endres2024can}. The model is configured with an input context window of 32{,}000 tokens and an output context window (maximum generation length) of 8{,}000 tokens. For APO, we use a separate customized instance of the same Gemini 2.0 model. The validation test suite is generated by the Gemini-cli AI agent \cite{geminicli}, which internally leverages the Gemini 2.5 Pro model.

\subsubsection{Benchmarks}
\label{sec:sec412}
We evaluate APRIL on three benchmark datasets comprising 81 synthesis tasks in total: 36 tasks from NumPy, 33 from scikit-learn, and 12 from SciPy. From these libraries, we additionally select 40 tasks to construct the training set used for APO and RLVR fine-tuning. The scikit-learn subset emphasizes canonical machine learning scenarios (e.g., GaussianProcessRegressor), in which the model is required to synthesize APIs that either transform input data or train a model on the provided data. The NumPy subset focuses on array-centric operations and numerically intensive routines, such as least-squares polynomial fitting. The SciPy subset primarily targets implementations of algebraic computations and classical statistical procedures, like Minkowski distance calculations.

\subsection{Results Analysis}
\label{sec:sec42}

\begin{table*}[h]
\centering
\caption{RQ1. Executability and test pass rate of APRIL}
\label{tab:table1}
\begin{tabular}{l|r|rr}
\toprule
\multicolumn{1}{c|}{Benchmark} & \#Tasks & \begin{tabular}[c]{@{}c@{}}  Executability\end{tabular} & \begin{tabular}[c]{@{}c@{}}   Test Pass Rate\end{tabular}  \\ \hline
NumPy & 36 & 36(100.0\%) & 35(97.2\%) \\
Scikit-learn & 33 & 33(100\%) & 30(90.9\%) \\ 
SciPy & 12 & 12(100.0\%) & 11(91.7\%) \\ \hline
Total & 81 & 81(100.0\%) & 76(93.8\%) \\
\bottomrule
\end{tabular}
\end{table*}

\subsubsection{RQ1}
\label{sec:sec421}
Our first evaluation of LLM-generated APIs examines effectiveness, defined by the proportion of tasks for which the synthesizer produces functionally correct APIs. We decompose this measure into two components: (i) execution rate and (ii) the rate of passing the validation test cases. The execution rate assesses whether the synthesized API can be built and executed without error. Because the LLM is provided with the target library and module context, and may therefore depend on in-library components, we evaluate executability by inserting the synthesized API into its designated module within the corresponding library and then attempting to build and execute the API. The second component measures whether the synthesized API passes the validation test suite associated with the task. When constructing the validation test suite for a given API, we instruct Gemini-cli to generate tests within the module’s existing test directory hierarchy (i.e., the directory designated for that module’s tests). We then compute the test success rate by executing the entire validation suite for the target API under the library’s test harness and recording the fraction of tests that pass.

Table~\ref{tab:table1} summarizes these metrics, reporting for each benchmark the number of tasks alongside two outcome measures: executability and test success rate. Empirically, all LLM-generated APIs built and executed successfully (100\% executability). Among these, 93.8\% passed their full validation suites. These results indicate high effectiveness in synthesizing functional APIs.

\begin{table*}[h]
\centering
\caption{RQ2. Comparison of APRIL with the baseline across three benchmarks}
\label{tab:table2}
\begin{tabular}{l|r|r|r}
\toprule
\multicolumn{1}{l|}{Benchmark} & \#Tasks & \begin{tabular}[c]{@{}c@{}}Success Rate: Baseline \end{tabular} & \begin{tabular}[c]{@{}c@{}}Success Rate: APRIL \end{tabular} \\ \hline
NumPy & 36 & 29(80.6\%) & 35(97.2\%) \\
Scikit-learn & 33 & 25(76.0\%) & 30(90.9\%) \\
SciPy & 12 & 9(75.0\%) & 11(91.7\%) \\ 
\hline
Total & 81 & 63(77.8\%) & 76(93.8\%) \\
\bottomrule
\end{tabular}
\end{table*}

\subsubsection{RQ2}
\label{sec:sec422}
We address RQ2 by comparing the success rate of APRIL against a baseline that performs synthesis using the initial prompt on Gemini 2.0 model. We execute all 81 tasks from the three benchmarks. The results are presented in Table~\ref{tab:table2}. The results suggest that APRIL can outperform the state-of-the-art LLMs baseline across all three benchmarks. On average, it achieves improvements of 16.6\%, 14.9\%, and 16.7\% over the baseline on the three benchmarks, respectively.

\begin{figure*}[h]
    \centering
    \includegraphics[width=0.86\textwidth,clip]{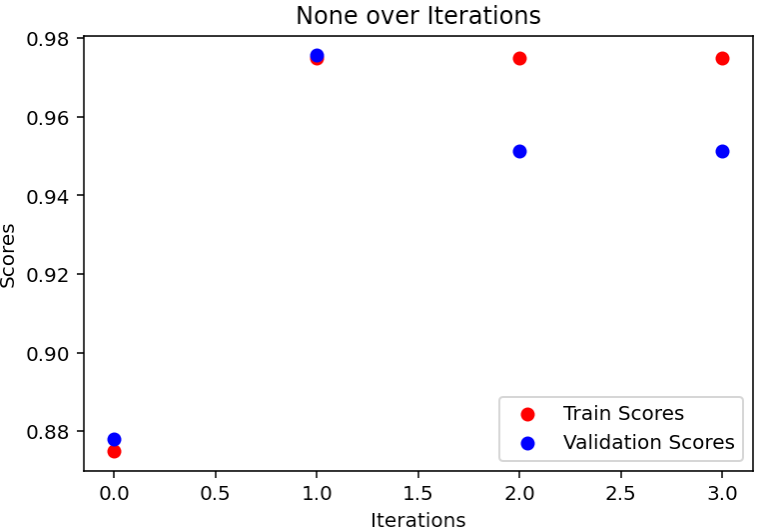}
    \caption{Potential iterative improvement from APO on RLVR fine-tuned Model}
    \label{fig:figure4}
\end{figure*}

Due to time and space constraints, we do not conduct an ablation study isolating the individual contributions of APO and RLVR. Moreover, applying APO to the RLVR–fine-tuned model suggests additional performance gains (Fig.~\ref{fig:figure4}); however, a comprehensive analysis of this follow-on step is beyond the scope of this paper.

\begin{table*}[h]
\centering
\caption{RQ3. Gemini-cli generated test counts and required iterations}
\label{tab:table3}
\begin{tabular}{l|r|r|r}
\toprule
\multicolumn{1}{l|}{Benchmark} & \#Tasks & \begin{tabular}[c]{@{}c@{}}Test count(Avg) \end{tabular} & \begin{tabular}[c]{@{}c@{}}Iterations count(Avg) \end{tabular} \\ \hline
NumPy & 36 & 258(7.2) & 86(2.4) \\
Scikit-learn & 33 & 282(8.5) & 70(2.1) \\
SciPy & 12 & 115(9.6) & 19(1.6) \\ 
\hline
Total & 81 & 655(8.1) & 175(2.2) \\
\bottomrule
\end{tabular}
\end{table*}

\subsubsection{RQ3}
\label{sec:sec423}
We assess the test-generation capability of Gemini-cli along two dimensions: (i) the number of validation tests produced per API and (ii) the number of quality-evaluation iterations required to obtain a comprehensive test suite for that API. As summarized in Table~\ref{tab:table3}, Gemini-cli generates, on average, 7.2 tests per API for NumPy, 8.5 for scikit-learn, and 9.6 for SciPy. The system requires an average of 2.2 iterations of the quality-evaluation loop to converge to a comprehensive test suite per API. These findings indicate that Gemini-cli is an effective tool for generating validation tests for scientific Python APIs, highlighting its potential to support automated program synthesis pipelines and rigorous evaluation workflows.

\section{Related Work}
\textbf{Program synthesis} has been a significant area of research, with various approaches proposed to generate programs efficiently. Traditional methods, such as Transit \cite{transit} and Escher \cite{escher}, rely on bottom-up enumerative synthesis, systematically exploring the program space to identify correct solutions. Brahma \cite{loopfree} introduces an efficient SMT-based encoding for synthesizing straight-line programs that involve multiple assignments to intermediate variables. Component-based synthesizers, such as SyPet \cite{sypet}, focus on generating Java programs from examples by leveraging arbitrary libraries. However, SyPet is limited to synthesizing sequences of method calls and does not support control structures like loops or conditionals. Similarly, Python-based synthesizers such as 
TFCoder \cite{tfcoder}, AutoPandas \cite{autopandas}, and Wrex \cite{wrex} primarily target one-liners or sequences of method calls. These tools also provide limited support for control structures, restricting their applicability for more complex program synthesis tasks. While these methods demonstrate the potential of automated program synthesis, their limitations in handling control structures underline the need for more versatile approaches.

\textbf{Synthesis with control structures.} EdSynth \cite{Edsynth} support program synthesis with control flow by lazily initializing candidates during the execution of provided tests. The execution of partially completed candidates determines the generation of future candidates, making EdSynth particularly effective for synthesis tasks that involve multiple API sequences in both the conditions and bodies of loops or branches.
FrAngel \cite{frangel} is another notable tool that supports component-based synthesis for Java programs with control structures. Unlike many traditional methods, FrAngel relies on function-level specifications and, in principle, does not require users to have a detailed understanding of the algorithm or intermediate variables. However, in practice, to make the synthesis process feasible, users must provide a high-quality test suite, covering base and corner cases. This requirement still necessitates some level of algorithmic knowledge, and FrAngel’s relatively slow performance makes it less suitable for interactive scenarios.
LooPy \cite{loopy}, introduces the concept of an Intermediate State Graph (ISG), which compactly represents a vast space of code snippets composed of multiple assignment statements and conditionals. By engaging human as an oracle to address incomplete parts of the loop, LooPy achieves a balance between automation and interactivity. Its ability to solve a wide range of synthesis tasks at interactive speeds makes it a practical tool for use cases requiring real-time feedback and adjustments.
These tools demonstrate progress in addressing the challenges of synthesis with control structures, but they also highlight trade-offs between usability, required user input, and performance.

\textbf{Large Language Models (LLMs)} have recently shown effectiveness in various software development tasks, including program synthesis\cite{jain2022jigsaw,hong2025effectiveness,zhong2025approach} and test generation\cite{xia2024fuzz4all,jiang2024generating}. By associating document text with code from a large training set, LLMs can generate program code from natural language prompts\cite{ugare2024improving,spiess2024quality}. Reusable API\cite{reuseableAPI} uses LLMs to generate APIs from code snippets collected from Stack Overflow and shows significant results in identifying API parameters and return types. However, they provide all components and dependencies to LLMs and only require LLMs to create new APIs using existing functions. Test-driven program synthesis remains an under-researched topic. How well do LLMs perform in generating entire APIs with just a few input/output examples that even end users can easily prepare? In this paper, we explore the application of LLMs in generating API implementations and finds that LLMs are effective in understanding test cases and generating viable APIs. We also show that LLMs are able to generate accurate APIs even with an incomplete set of test cases, which is very convenient. To the best of our knowledge, our work is the first systematic study of using LLMs with auto prompt engineering and fine-tuning to synthesize complex APIs.


\section{Conclusion}
We propose integrating fine-tuned large language models (LLMs) with Automatic Prompt Optimization (APO) for API synthesis. LLMs capture developer intent and context, enabling more effective synthesis. Empirically, our method outperforms a strong baseline that uses a state-of-the-art LLM with a manually engineered prompt. We also assess Gemini-cli’s ability to generate comprehensive validation test suites, underscoring its potential to support automated program synthesis and rigorous evaluation. The approach requires minimal user interaction, simplifying synthesis. Overall, the results support LLMs as a practical foundation for complex API synthesis.

\bibliographystyle{abbrv}
\bibliography{main}

\begin{thebibliography}{10}

\bibitem{geminicli}
Gemini cli, 2025.
\newblock \url{https://cloud.google.com/gemini/docs/codeassist/gemini-cli}.

\bibitem{escher}
A.~Albarghouthi, S.~Gulwani, and Z.~Kincaid.
\newblock Recursive program synthesis.
\newblock In {\em Computer Aided Verification: 25th International Conference, CAV 2013, Saint Petersburg, Russia, July 13-19, 2013. Proceedings 25}, pages 934--950. Springer, 2013.

\bibitem{autopandas}
R.~Bavishi, C.~Lemieux, R.~Fox, K.~Sen, and I.~Stoica.
\newblock Autopandas: neural-backed generators for program synthesis.
\newblock {\em Proc. ACM Program. Lang.}, 3(OOPSLA), Oct. 2019.

\bibitem{wrex}
I.~Drosos, T.~Barik, P.~J. Guo, R.~DeLine, and S.~Gulwani.
\newblock Wrex: A unified programming-by-example interaction for synthesizing readable code for data scientists.
\newblock In {\em Proceedings of the 2020 CHI Conference on Human Factors in Computing Systems}, CHI '20, page 1–12, New York, NY, USA, 2020. Association for Computing Machinery.

\bibitem{endres2024can}
M.~Endres, S.~Fakhoury, S.~Chakraborty, and S.~K. Lahiri.
\newblock Can large language models transform natural language intent into formal method postconditions?
\newblock {\em Proceedings of the ACM on Software Engineering}, 1(FSE):1889--1912, 2024.

\bibitem{syntable}
Y.~Feng, R.~Martins, J.~Van~Geffen, I.~Dillig, and S.~Chaudhuri.
\newblock Component-based synthesis of table consolidation and transformation tasks from examples.
\newblock In {\em Proceedings of the 38th ACM SIGPLAN Conference on Programming Language Design and Implementation}, PLDI 2017, page 422–436, New York, NY, USA, 2017. Association for Computing Machinery.

\bibitem{sypet}
Y.~Feng, R.~Martins, Y.~Wang, I.~Dillig, and T.~W. Reps.
\newblock Component-based synthesis for complex apis.
\newblock In {\em Proceedings of the 44th ACM SIGPLAN Symposium on Principles of Programming Languages}, POPL '17, page 599–612, New York, NY, USA, 2017. Association for Computing Machinery.

\bibitem{loopy}
K.~Ferdowsifard, S.~Barke, H.~Peleg, S.~Lerner, and N.~Polikarpova.
\newblock Loopy: interactive program synthesis with control structures.
\newblock {\em Proc. ACM Program. Lang.}, 5(OOPSLA), Oct. 2021.

\bibitem{syndstrans}
J.~K. Feser, S.~Chaudhuri, and I.~Dillig.
\newblock Synthesizing data structure transformations from input-output examples.
\newblock {\em SIGPLAN Not.}, 50(6):229–239, June 2015.

\bibitem{stringio}
S.~Gulwani.
\newblock Automating string processing in spreadsheets using input-output examples.
\newblock {\em SIGPLAN Not.}, 46(1):317–330, Jan. 2011.

\bibitem{10.1145/1993316.1993506}
S.~Gulwani, S.~Jha, A.~Tiwari, and R.~Venkatesan.
\newblock Synthesis of loop-free programs.
\newblock {\em SIGPLAN Not.}, 46(6):62–73, June 2011.

\bibitem{loopfree}
S.~Gulwani, S.~Jha, A.~Tiwari, and R.~Venkatesan.
\newblock Synthesis of loop-free programs.
\newblock In {\em Proceedings of the 32nd ACM SIGPLAN Conference on Programming Language Design and Implementation}, PLDI '11, page 62–73, New York, NY, USA, 2011. Association for Computing Machinery.

\bibitem{mimic}
S.~Heule, M.~Sridharan, and S.~Chandra.
\newblock Mimic: computing models for opaque code.
\newblock In {\em Proceedings of the 2015 10th Joint Meeting on Foundations of Software Engineering}, ESEC/FSE 2015, page 710–720, New York, NY, USA, 2015. Association for Computing Machinery.

\bibitem{hong2025effectiveness}
Y.~Hong, S.~Jiang, Y.~Fu, and S.~Khurshid.
\newblock On the effectiveness of large language models in writing alloy formulas.
\newblock {\em arXiv preprint arXiv:2502.15441}, 2025.

\bibitem{jain2022jigsaw}
N.~Jain, S.~Vaidyanath, A.~Iyer, N.~Natarajan, S.~Parthasarathy, S.~Rajamani, and R.~Sharma.
\newblock Jigsaw: Large language models meet program synthesis.
\newblock In {\em Proceedings of the 44th International Conference on Software Engineering}, pages 1219--1231, 2022.

\bibitem{oguide}
S.~Jha, S.~Gulwani, S.~A. Seshia, and A.~Tiwari.
\newblock Oracle-guided component-based program synthesis.
\newblock In {\em Proceedings of the 32nd ACM/IEEE International Conference on Software Engineering - Volume 1}, ICSE '10, page 215–224, New York, NY, USA, 2010. Association for Computing Machinery.

\bibitem{jiang2025cascade}
S.~Jiang, P.~Kovuri, D.~Tao, and Z.~Tan.
\newblock Cascade: Llm-powered javascript deobfuscator at google.
\newblock {\em arXiv preprint arXiv:2507.17691}, 2025.

\bibitem{jiang2024generating}
S.~Jiang, C.~Zhu, and S.~Khurshid.
\newblock Generating executable oracles to check conformance of client code to requirements of jdk javadocs using llms.
\newblock {\em arXiv preprint arXiv:2411.01789}, 2024.

\bibitem{lambert2025tulu}
N.~Lambert, J.~Morrison, V.~Pyatkin, S.~Huang, H.~Ivison, F.~Brahman, L.~J.~V. Miranda, A.~Liu, N.~Dziri, X.~Lyu, Y.~Gu, S.~Malik, V.~Graf, J.~D. Hwang, J.~Yang, R.~L. Bras, O.~Tafjord, C.~Wilhelm, L.~Soldaini, N.~A. Smith, Y.~Wang, P.~Dasigi, and H.~Hajishirzi.
\newblock Tulu 3: Pushing frontiers in open language model post-training.
\newblock In {\em Second Conference on Language Modeling}, 2025.

\bibitem{lamothe2021systematic}
M.~Lamothe, Y.-G. Gu\'{e}h\'{e}neuc, and W.~Shang.
\newblock A systematic review of api evolution literature.
\newblock {\em ACM Comput. Surv.}, 54(8), Oct. 2021.

\bibitem{reuseableAPI}
Y.~Mai, Z.~Gao, X.~Hu, L.~Bao, Y.~Liu, and J.~Sun.
\newblock Are human rules necessary? generating reusable apis with cot reasoning and in-context learning.
\newblock {\em Proc. ACM Softw. Eng.}, 1(FSE), July 2024.

\bibitem{jungloid}
D.~Mandelin, L.~Xu, R.~Bod\'{\i}k, and D.~Kimelman.
\newblock Jungloid mining: helping to navigate the api jungle.
\newblock In {\em Proceedings of the 2005 ACM SIGPLAN Conference on Programming Language Design and Implementation}, PLDI '05, page 48–61, New York, NY, USA, 2005. Association for Computing Machinery.

\bibitem{icmlpbe}
A.~K. Menon, O.~Tamuz, S.~Gulwani, B.~Lampson, and A.~T. Kalai.
\newblock A machine learning framework for programming by example.
\newblock In {\em Proceedings of the 30th International Conference on International Conference on Machine Learning - Volume 28}, ICML'13, page I–187–I–195. JMLR.org, 2013.

\bibitem{nam2024using}
D.~Nam, A.~Macvean, V.~Hellendoorn, B.~Vasilescu, and B.~Myers.
\newblock Using an llm to help with code understanding.
\newblock In {\em Proceedings of the IEEE/ACM 46th International Conference on Software Engineering}, pages 1--13, 2024.

\bibitem{nikolov2025google}
S.~Nikolov, D.~Codecasa, A.~Sjovall, M.~Tabachnyk, S.~Taneja, C.~Ziftci, and S.~Chandra.
\newblock How is google using ai for internal code migrations?
\newblock In {\em 2025 IEEE/ACM 47th International Conference on Software Engineering (ICSE)}. IEEE, 2025.

\bibitem{testdriven}
D.~Perelman, S.~Gulwani, D.~Grossman, and P.~Provost.
\newblock Test-driven synthesis.
\newblock In {\em Proceedings of the 35th ACM SIGPLAN Conference on Programming Language Design and Implementation}, PLDI '14, page 408–418, New York, NY, USA, 2014. Association for Computing Machinery.

\bibitem{pryzantetal2023automatic}
R.~Pryzant, D.~Iter, J.~Li, Y.~Lee, C.~Zhu, and M.~Zeng.
\newblock Automatic prompt optimization with ``gradient descent'' and beam search.
\newblock In H.~Bouamor, J.~Pino, and K.~Bali, editors, {\em Proceedings of the 2023 Conference on Empirical Methods in Natural Language Processing}, pages 7957--7968, Singapore, Dec. 2023. Association for Computational Linguistics.

\bibitem{schaeffer2024emergent}
R.~Schaeffer, B.~Miranda, and S.~Koyejo.
\newblock Are emergent abilities of large language models a mirage?
\newblock {\em Advances in Neural Information Processing Systems}, 36, 2024.

\bibitem{tfcoder}
K.~Shi, D.~Bieber, and R.~Singh.
\newblock Tf-coder: Program synthesis for tensor manipulations.
\newblock {\em ACM Trans. Program. Lang. Syst.}, 44(2), May 2022.

\bibitem{frangel}
K.~Shi, J.~Steinhardt, and P.~Liang.
\newblock Frangel: component-based synthesis with control structures.
\newblock {\em Proc. ACM Program. Lang.}, 3(POPL), Jan. 2019.

\bibitem{spiess2024quality}
C.~Spiess, D.~Gros, K.~S. Pai, M.~Pradel, M.~R.~I. Rabin, S.~Jha, P.~Devanbu, and T.~Ahmed.
\newblock Quality and trust in llm-generated code.
\newblock {\em arXiv preprint arXiv:2402.02047}, 2024.

\bibitem{proveri}
S.~Srivastava, S.~Gulwani, and J.~S. Foster.
\newblock From program verification to program synthesis.
\newblock {\em SIGPLAN Not.}, 45(1):313–326, Jan. 2010.

\bibitem{transit}
A.~Udupa, A.~Raghavan, J.~V. Deshmukh, S.~Mador-Haim, M.~M. Martin, and R.~Alur.
\newblock Transit: specifying protocols with concolic snippets.
\newblock In {\em Proceedings of the 34th ACM SIGPLAN Conference on Programming Language Design and Implementation}, PLDI '13, page 287–296, New York, NY, USA, 2013. Association for Computing Machinery.

\bibitem{ugare2024improving}
S.~Ugare, T.~Suresh, H.~Kang, S.~Misailovic, and G.~Singh.
\newblock Improving llm code generation with grammar augmentation.
\newblock {\em arXiv preprint arXiv:2403.01632}, 2024.

\bibitem{xia2024fuzz4all}
C.~S. Xia, M.~Paltenghi, J.~Le~Tian, M.~Pradel, and L.~Zhang.
\newblock Fuzz4all: Universal fuzzing with large language models.
\newblock In {\em Proceedings of the IEEE/ACM 46th International Conference on Software Engineering}, pages 1--13, 2024.

\bibitem{Edsynth}
Z.~Yang, J.~Hua, K.~Wang, and S.~Khurshid.
\newblock Edsynth: Synthesizing api sequences with conditionals and loops.
\newblock In {\em 2018 IEEE 11th International Conference on Software Testing, Verification and Validation (ICST)}, pages 161--171, 2018.

\bibitem{zhong2025approach}
H.~Zhong, S.~Jiang, and S.~Khurshid.
\newblock An approach for api synthesis using large language models.
\newblock {\em arXiv preprint arXiv:2502.15246}, 2025.

\end{thebibliography}
\end{document}